\documentclass[a4paper,12pt]{article}
\usepackage{amsfonts}
\usepackage{amsmath}
\usepackage{graphicx}

\numberwithin{equation}{section}

\begin{document}

\begin{center}
{\LARGE \textbf{View on N-dimensional spherical harmonics from the quantum
mechanical P\"{o}schl-Teller potential well}}\vspace{0.5cm}

{\large A. Smirnov}\footnote{%
email: smirnov@ufs.br; smirnov.globe@gmail.com}

\textit{Universidade Federal de Sergipe, S\~{a}o Crist\'{o}v\~{a}o, Brazil}%
\vspace{0.5cm}

\textbf{Abstract}
\end{center}

{\small In this work we propose an approach of obtaining of N-dimensional
spherical harmonics based exclusively on the methods of solutions of
differential equations and the use of the special functions properties. We
deduce the Laplace-Beltrami operator on the N-sphere, indicate some
instructive relations for the metric, and demonstrate the procedure of
separation of the variables. We show that the ordinary differential
equations for every variable, except one, can be reduced to the Schr\"{o}%
dinger equation (SE) with the symmetric P\"{o}schl-Teller (SPT) potential
well by means of certain substitutions. We also exhibit that the solutions
of SE with SPT potential are expressed in terms of the Gegenbauer
polynomials. The eigenvalues of the Laplace-Beltrami operator and the
characteristic numbers of the spherical harmonics are obtained with the use
of the properties of the spectrum of SE with SPT potential. The spherical
harmonics are constructed as a product of the eigenfunctions of SE with SPT
potential multiplied by a easily computable factor function and expressed in
terms of the Gegenbauer polynomials. The work has a pedagogical character to
some extent.}\vspace{0.5cm}

\section{Introduction}

In this paper we propose an approach of obtaining of N-dimensional spherical
harmonics based on the method of separation of variables and on the use of
the properties of the special functions. We point out the works which are
most substantial on the question in our opinion. The first thorough
description of obtaining of N-dimensional spherical harmonics was apparently
made in the monograph \cite{53be} where the harmonic polynomials theory is
applied and the generating function of Gegenbauer polynomials is utilized.
In \cite{53be} the spherical harmonics are obtained in terms of Gegenbauer
polynomials. The theorems of orthogonality and completeness of the system of
the spherical harmonics are proved, the addition theorem and the other
important relations are shown. In the work \cite{60louck} a generalized
angular momentum technique is applied. In particular, the eigenvalues of a
square of the orbital angular momentum operator in N dimensions are
determined. The eigenvectors of this operator can be interpreted as
N-dimensional spherical harmonics. In the paper \cite{87h} scalar spherical
harmonics in N dimensions are obtained as a starting point for discussion of
symmetric tensor spherical harmonics in N dimensions. The scalar spherical
harmonics are obtained in terms of the associated Legendre functions for the
purposes of the work. The necessary properties of the spherical harmonics
are demonstrated and eigenvalues of the Laplace-Beltrami operator on
N-dimensional sphere are determined. In the paper \cite{10a} a construction
of N-dimensional spherical harmonics is fulfilled by means of a harmonic
projection. This method is proposed as complementary to the methods of the
group theory. In addition we mention a recent pedagogical work \cite{14ef}
in which N-dimensional spherical geometry is discussed with a description of
the use of Legendre polynomials. In our paper we propose an approach which
is distinct from the others described in the accessible literature.

We also note that papers in which the spherical harmonics of various
dimensions are utilized in one or another mode for discussions of diverse
problems have continued to appear. We mention a few recent papers of this
kind. For example, in the paper \cite{14asmr} the scalar harmonics on a four
sphere are analyzed using a associated Legendre function. Then, they are
used for construction of two types of vector harmonics and three types of
tensor harmonics on a four sphere. In \cite{13gmg} the spherical harmonics
in N dimensions are used for the construction of summation by parts
finite-differencing numerical methods of solution of wave equations in N+2
spacetime dimensions. In \cite{16ahqr} the explicit vector spherical
harmonics on the 3-sphere are constructed with the use of the technique of
p-forms. In \cite{16f} the spherical harmonics on 2-sphere are considered
from the viewpoint of su(2) Lie algebra symmetry realized in quantization of
the magnitude and z-component of angular momentum operator in terms of the
azimuthal and magnetic quantum numbers. The azimuthal quantum number is
associated with an additional ladder symmetry. In \cite{16f} it is shown
that simultaneous realization of these two symmetries inherits the positive
and negative integer discrete irreducible representations for su(1,1) Lie
algebra via the spherical harmonics on the sphere as a compact manifold.

The paper is organized as follows. In Sec. 2 we introduce notations and
obtain the Laplace-Beltrami operator on the N-sphere. We reproduce principal
steps of computing of the metric and indicate its properties necessary for
application of the method of separation of variables. In Sec. 3 we apply the
method of separation of variables and obtain a set of the ordinary
differential equations of a certain form. In Sec. 4 we describe the
solutions of the Schr\"{o}dinger equation with the symmetric P\"{o}%
schl-Teller potential and present them in terms of the Gegenbauer
polynomials. In Sec. 5 we solve the set of the ordinary differential
equations obtained in Sec. 3 applying the solutions of Sec. 4. In Sec. 6 we
construct the N-dimensional spherical harmonics and make some comments. In
Sec. 7 we present conclusions of the work.

\section{Laplace-Beltrami operator on the N-sphere}

To construct N-dimensional spherical harmonics we solve the Laplace equation
in spherical coordinates in $N+1$ dimensions. To this end we use a technique
of the Laplace-Beltrami operator. We reproduce some details of computation
of the metric and the Laplace-Beltrami operator, mainly for didactical
purposes. We use the spherical coordinates in $N+1$ dimensions $\left(
r,\theta _{1},\theta _{2},...,\theta _{N}\right) $ in the following form:%
\begin{eqnarray}
&&x_{N+1}=r\cos \theta _{N}\ ,  \label{r1a} \\
&&x_{N}=r\sin \theta _{N}\cos \theta _{N-1}\ ,  \notag \\
&&x_{N-1}=r\sin \theta _{N}\sin \theta _{N-1}\cos \theta _{N-2}\ ,  \notag \\
&&...  \notag \\
&&x_{3}=r\sin \theta _{N}\sin \theta _{N-1}...\sin \theta _{3}\cos \theta
_{2}\ ,  \notag \\
&&x_{2}=r\sin \theta _{N}\sin \theta _{N-1}...\sin \theta _{3}\sin \theta
_{2}\cos \theta _{1}\ ,  \notag \\
&&x_{1}=r\sin \theta _{N}\sin \theta _{N-1}...\sin \theta _{3}\sin \theta
_{2}\sin \theta _{1}  \notag
\end{eqnarray}%
with the ranges:%
\begin{equation*}
0\leq r<\infty ,\ 0\leq \theta _{1}<2\pi ,\ 0\leq \theta _{i}\leq \pi ,\
i=2,3,...,N\ 
\end{equation*}%
so that $x_{1}^{2}+x_{2}^{2}+...+x_{N+1}^{2}=r^{2}.$ To obtain the Laplacian
we use the Laplace-Beltrami operator applied to a scalar function $u:$%
\begin{equation}
\nabla ^{2}u=\frac{1}{\sqrt{\left\vert g\right\vert }}\partial _{i}\left( 
\sqrt{\left\vert g\right\vert }g^{ij}\partial _{j}u\right)  \label{lbo}
\end{equation}%
where $g^{ij}=g_{ij}^{-1}$ is the inverse of the metric tensor $g_{ij},$ $g=$%
det$\left( g_{ij}\right) $ is the determinant of $g_{ij},$ $\partial _{i}$
is the derivative with respect to the variable $q_{i}$: $q_{N+1}=r,\
q_{i}=\theta _{i},\ i=1,2,...,N,$ and summation over $i,\ j$ is implied.

Using the relation of the spherical coordinates with the Cartesian
coordinates (\ref{r1a}) one can determine the metric $g_{ij}$ as follows:%
\begin{equation}
g_{ij}=\sum_{k=1}^{N+1}\frac{\partial x_{k}}{\partial q_{i}}\frac{\partial
x_{k}}{\partial q_{j}}\ .  \label{c04}
\end{equation}%
For the derivatives in (\ref{c04}) one gets:%
\begin{equation}
\frac{\partial x_{k}}{\partial q_{N+1}}=\frac{x_{k}}{r},\ \frac{\partial
x_{k}}{\partial q_{i}}=\left\{ 
\begin{array}{l}
0,\ k\geq i+2\ , \\ 
-x_{i+1}\tan \theta _{i},\ k=i+1\ , \\ 
x_{k}\cot \theta _{i},\ k\leq i\ .%
\end{array}%
\right.  \label{c05}
\end{equation}%
We also note that%
\begin{equation}
\sum_{k=1}^{N+1}x_{k}^{2}=r^{2},\ \sum_{k=1}^{i}x_{k}^{2}=x_{i+1}^{2}\tan
^{2}\theta _{i}\ .  \label{c06}
\end{equation}%
Then with the use of Eqs. (\ref{c05}), (\ref{c06}) one gets the components
of the metric:%
\begin{equation*}
g_{N+1,N+1}=\sum_{k=1}^{N+1}\left( \frac{\partial x_{k}}{\partial r}\right)
^{2}=\sum_{k=1}^{N+1}\frac{x_{k}^{2}}{r^{2}}=1\ ,
\end{equation*}%
\begin{eqnarray*}
&&g_{ii}=\sum_{k=1}^{N+1}\left( \frac{\partial x_{k}}{\partial q_{i}}\right)
^{2}=\sum_{k=1}^{i}\left( \frac{\partial x_{k}}{\partial q_{i}}\right)
^{2}+\left( \frac{\partial x_{i+1}}{\partial q_{i}}\right)
^{2}+\sum_{k=i+2}^{N+1}\left( \frac{\partial x_{k}}{\partial q_{i}}\right)
^{2} \\
&&\left. =\right. \sum_{k=1}^{i}x_{k}^{2}\cot ^{2}\theta
_{i}+x_{i+1}^{2}\tan ^{2}\theta _{i}=x_{i+1}^{2}\left( 1+\tan ^{2}\theta
_{i}\right) =\frac{x_{i+1}^{2}}{\cos ^{2}\theta _{i}},\ i\leq N\ ,
\end{eqnarray*}%
\begin{eqnarray*}
&&g_{ij}=\sum_{k=1}^{N+1}\frac{\partial x_{k}}{\partial q_{i}}\frac{\partial
x_{k}}{\partial q_{j}}=\sum_{k=1}^{i}\frac{\partial x_{k}}{\partial q_{i}}%
\frac{\partial x_{k}}{\partial q_{j}}+\frac{\partial x_{i+1}}{\partial q_{i}}%
\frac{\partial x_{i+1}}{\partial q_{j}}+\sum_{k=i+2}^{N+1}\frac{\partial
x_{k}}{\partial q_{i}}\frac{\partial x_{k}}{\partial q_{j}} \\
&&\left. =\right. \sum_{k=1}^{i}x_{k}^{2}\cot \theta _{i}\cot \theta
_{j}-x_{i+1}^{2}\tan \theta _{i}\cot \theta _{j}=x_{i+1}^{2}\tan \theta
_{i}\cot \theta _{j}-x_{i+1}^{2}\tan \theta _{i}\cot \theta _{j}=0,\ i<j
\end{eqnarray*}%
therefore%
\begin{equation}
g_{ij}=\text{diag}\left( \eta _{i}\right) ,\ \eta _{N+1}=1,\ \eta _{i}=\frac{%
x_{i+1}^{2}}{\cos ^{2}\theta _{i}}=r^{2}\prod_{k=i+1}^{N}\sin ^{2}\theta
_{k},\ 1\leq i\leq N\ .  \label{c07}
\end{equation}%
For the determinant $g$ one gets:%
\begin{equation*}
g=\det g_{ij}=\prod_{i=1}^{N+1}g_{ii}=\prod_{i=1}^{N}\frac{x_{i+1}^{2}}{\cos
^{2}\theta _{i}}=r^{2N}\prod_{k=2}^{N}\sin ^{2\left( k-1\right) }\theta _{k}
\end{equation*}%
and respectively%
\begin{equation}
\sqrt{\left\vert g\right\vert }=\prod_{i=1}^{N}\frac{x_{i+1}}{\cos \theta
_{i}}=r^{N}\prod_{k=2}^{N}\sin ^{k-1}\theta _{k}\ .  \label{c08}
\end{equation}%
For the inverse tensor $g^{ij}$ one has: 
\begin{equation}
g^{ij}=\left( g_{ij}\right) ^{-1}=\text{diag}\left( h^{i}\right) ,\
h^{N+1}=1,\ h^{i}=\eta _{i}^{-1}=r^{-2}\prod_{k=i+1}^{N}\sin ^{-2}\theta
_{k},\ 1\leq i\leq N\ .  \label{c09}
\end{equation}%
Now we compute the Laplace-Beltrami operator: 
\begin{equation}
\nabla ^{2}u=\frac{1}{\sqrt{\left\vert g\right\vert }}\partial _{i}\left( 
\sqrt{\left\vert g\right\vert }g^{ij}\partial _{j}u\right) =\frac{1}{\sqrt{%
\left\vert g\right\vert }}\left( \partial _{i}\sqrt{\left\vert g\right\vert }%
\right) g^{ij}\partial _{j}u+\left( \partial _{i}g^{ij}\right) \partial
_{j}u+g^{ij}\partial _{i}\partial _{j}u  \label{c10}
\end{equation}%
where we imply that the derivatives inside the brackets on the rhs are
applied only to the quantities inside the brackets. For the first term of
Eq. (\ref{c10}) one can see that%
\begin{equation*}
\frac{1}{\sqrt{\left\vert g\right\vert }}\left( \partial _{N+1}\sqrt{%
\left\vert g\right\vert }\right) =Nr^{-1},\ \partial _{1}\sqrt{\left\vert
g\right\vert }=0,\ \frac{1}{\sqrt{\left\vert g\right\vert }}\left( \partial
_{i}\sqrt{\left\vert g\right\vert }\right) =\left( i-1\right) \cot \theta
_{i},\ 2\leq i\leq N
\end{equation*}%
then%
\begin{equation*}
\frac{1}{\sqrt{\left\vert g\right\vert }}\left( \partial _{i}\sqrt{%
\left\vert g\right\vert }\right) g^{ij}\partial
_{j}u=\sum_{i=1}^{N}h^{i}\left( i-1\right) \cot \theta _{i}\partial
_{i}u+Nr^{-1}\partial _{r}u\ .
\end{equation*}%
For the second term one can see that $\partial _{i}g^{ij}=0,\ \forall j$ (no
summation) then 
\begin{equation*}
\left( \partial _{i}g^{ij}\right) \partial _{j}u=0\ .
\end{equation*}%
For the third term one has:%
\begin{equation*}
g^{ij}\partial _{i}\partial _{j}u=\sum_{i=1}^{N+1}h^{i}\partial
_{i}^{2}u=\sum_{i=1}^{N}h^{i}\partial _{i}^{2}u+\partial
_{N+1}^{2}u=\sum_{i=1}^{N}h^{i}\partial _{i}^{2}u+\partial _{r}^{2}u\ .
\end{equation*}%
Therefore%
\begin{equation*}
\nabla ^{2}u=\sum_{i=1}^{N}h^{i}\left[ \left( i-1\right) \cot \theta
_{i}\partial _{i}u+\partial _{i}^{2}u\right] +r^{-N}\partial _{r}\left(
r^{N}\partial _{r}u\right) \ .
\end{equation*}%
It is convenient to introduce the quantities%
\begin{equation}
h_{\left( N\right) }^{i}=r^{2}h^{i},\ 1\leq i\leq N\ ,  \label{c14}
\end{equation}%
which depend only on the angular variables $\theta _{i}$ and does not depend
on the variable $r.$ Then we arrive at the final form of the Laplace
operator in the spherical coordinates:%
\begin{equation}
\nabla ^{2}u=r^{-2}\nabla _{S^{N}}^{2}u+r^{-N}\partial _{r}\left(
r^{N}\partial _{r}u\right)  \label{c11}
\end{equation}%
where%
\begin{equation}
\nabla _{S^{N}}^{2}u=\sum_{i=1}^{N}h_{\left( N\right) }^{i}\left[ \left(
i-1\right) \cot \theta _{i}\partial _{i}u+\partial _{i}^{2}u\right]
\label{c12}
\end{equation}%
is the Laplace-Beltrami operator on the unit N-sphere $S^{N}.$

\section{Separation of variables}

In order to solve the Laplace equation%
\begin{equation*}
\nabla ^{2}u=0
\end{equation*}%
we present $u$ as%
\begin{equation*}
u\left( r,\theta _{1},\theta _{2},...,\theta _{N}\right) =R\left( r\right)
W_{N}\left( \theta _{1},\theta _{2},...,\theta _{N}\right)
\end{equation*}%
and we come to%
\begin{equation*}
\frac{1}{R}r^{-N}\partial _{r}\left( r^{N}\partial _{r}R\right) =-\frac{1}{%
W_{N}}\Delta _{S^{N}}W_{N}=\lambda _{N}=const
\end{equation*}%
because the lhs of the equality depends only on the radial variable $r$ and
the rhs depends only on the angular variables $\theta _{i}$, $\lambda _{N}$
is a separation constant.

We concentrate our attention on the equation:%
\begin{equation}
\nabla _{S^{N}}^{2}W_{N}+\lambda _{N}W_{N}=0  \label{c13}
\end{equation}%
which is the main point our work and whose solution represents the
N-dimensional spherical harmonics. It is useful to introduce the quantities%
\begin{equation}
h_{\left( k-1\right) }^{i}=\sin ^{2}\theta _{k}h_{\left( k\right) }^{i},\
1\leq i\leq k-1,\ 2\leq k\leq N\ .  \label{c15}
\end{equation}%
For a given value $k$ the set $\left\{ h_{\left( k\right) }^{i}\right\} $
depend only on the variables $\left( \theta _{1},...,\theta _{k}\right) $
and $h_{\left( k\right) }^{k}=1.$ Then we can write Eq. (\ref{c13}) as%
\begin{equation*}
\sum_{i=1}^{N-1}\sin ^{-2}\theta _{N}h_{\left( N-1\right) }^{i}\left[ \left(
i-1\right) \cot \theta _{i}\partial _{i}W_{N}+\partial _{i}^{2}W_{N}\right]
+h_{\left( N\right) }^{N}\left( \left( N-1\right) \cot \theta _{N}\partial
_{N}W_{N}+\partial _{N}^{2}W_{N}\right) =-\lambda _{N}W_{N}
\end{equation*}%
or in a compact form:%
\begin{equation}
\sin ^{-2}\theta _{N}\nabla _{S^{N-1}}^{2}W_{N}+\left( \left( N-1\right)
\cot \theta _{N}\partial _{N}W_{N}+\partial _{N}^{2}W_{N}\right) =-\lambda
_{N}W_{N}  \label{c16}
\end{equation}%
where the notation of Eq. (\ref{c12}) was utilized. We present $W_{N}$ as
follows:%
\begin{equation}
W_{N}\left( \theta _{1},\theta _{2},...,\theta _{N}\right) =W_{N-1}\left(
\theta _{1},\theta _{2},...,\theta _{N-1}\right) y_{N}\left( \theta
_{N}\right)  \label{c17}
\end{equation}%
then we come to 
\begin{equation*}
-\frac{1}{W_{N-1}}\Delta _{S^{N-1}}W_{N-1}=\frac{\sin ^{2}\theta _{N}}{y_{N}}%
\left( \left( N-1\right) \cot \theta _{N}\partial _{N}y_{N}+\partial
_{N}^{2}y_{N}\right) +\lambda _{N}\sin ^{2}\theta _{N}=const=\lambda _{N-1}
\end{equation*}%
because the lhs of the equality depends only on the variables $\left( \theta
_{1},...,\theta _{N-1}\right) $ and the rhs depends only on the variable $%
\theta _{N}$, $\lambda _{N-1}$ is a new separation constant. Thus we get the
equation for $W_{N-1}:$%
\begin{equation*}
\nabla _{S^{N-1}}^{2}W_{N-1}=-\lambda _{N-1}W_{N-1}
\end{equation*}%
and the equation for $y_{N}\left( \theta _{N}\right) $:%
\begin{equation*}
\left( \left( N-1\right) \cot \theta _{N}\partial _{N}y_{N}+\partial
_{N}^{2}y_{N}\right) +\left( \lambda _{N}-\frac{\lambda _{N-1}}{\sin
^{2}\theta _{N}}\right) y_{N}=0\ .
\end{equation*}%
Repeating this process we obtain the set of equations for $W_{k}=W_{k}\left(
\theta _{1},\theta _{2},...,\theta _{k}\right) :$%
\begin{equation}
\nabla _{S^{k}}^{2}W_{k}+\lambda _{k}W_{k}=0,\ 1\leq k\leq N  \label{hh.39}
\end{equation}%
and for $y_{k}=y_{k}\left( \theta _{k}\right) :$ 
\begin{equation}
\partial _{k}^{2}y_{k}+\left( k-1\right) \cot \theta _{k}\partial
_{k}y_{k}+\left( \lambda _{k}-\frac{\lambda _{k-1}}{\sin ^{2}\theta _{k}}%
\right) y_{k}=0,\ 2\leq k\leq N  \label{hh.r1}
\end{equation}%
with the set of $N$ separation constants $\lambda _{k}.$ For $k=1$ we get:%
\begin{equation}
\nabla _{S^{1}}^{2}y_{1}=\partial _{1}^{2}y_{1}=-\lambda _{1}y_{1}\ .
\label{hh.31}
\end{equation}%
The spherical harmonics $W_{N}$ are finally expressed in the form of a
product of $y_{k}\left( \theta _{k}\right) :$%
\begin{equation}
W_{N}\left( \theta _{1},\theta _{2},...,\theta _{N}\right) =y_{1}\left(
\theta _{1}\right) y_{2}\left( \theta _{2}\right) ...y_{N-1}\left( \theta
_{N-1}\right) y_{N}\left( \theta _{N}\right) \ .  \label{c01}
\end{equation}

\section{Symmetric P\"{o}schl-Teller potential}

Now we describe the solutions of the Schr\"{o}dinger equation with the
symmetric P\"{o}schl-Teller potential which will be used later in searching
for solutions of the equations (\ref{hh.r1}). We follow the exposition of
Ref. \cite{99f} (problem 38). In accordance with \cite{99f} the solutions of
the Schr\"{o}dinger equation with the P\"{o}schl-Teller potential of a
general form:%
\begin{equation}
-\frac{d^{2}\psi }{dx^{2}}+V_{PT}\left( x\right) \psi =q^{2}\psi ,\
V_{PT}\left( x\right) =c^{2}\left( \frac{\mu \left( \mu -1\right) }{\sin
^{2}\left( cx\right) }+\frac{\kappa \left( \kappa -1\right) }{\cos
^{2}\left( cx\right) }\right) ,\ 0\leq x\leq \frac{\pi }{2c};\ \mu >1,\
\kappa >1  \label{c18}
\end{equation}%
are given by the functions:%
\begin{equation}
\psi _{n}\left( x\right) =\sin ^{\mu }\left( cx\right) \cos ^{\kappa }\left(
cx\right) \left. _{2}F_{1}\right. \left( -n,n+\mu +\kappa ,\mu +\frac{1}{2}%
;\sin ^{2}\left( cx\right) \right)  \label{c19}
\end{equation}%
with the spectrum:%
\begin{equation}
q_{n}^{2}=c^{2}\left( 2n+\mu +\kappa \right) ^{2},\ n=0,1,2,...  \label{c20}
\end{equation}%
where $\left. _{2}F_{1}\right. \left( a,b,c;x\right) $ is the hypergeometric
function. In the case of the symmetric P\"{o}schl-Teller potential we have $%
\mu =\kappa $ and we can write the potential in the form:%
\begin{equation*}
V_{SPT}\left( x\right) =4c^{2}\frac{\mu \left( \mu -1\right) }{\sin
^{2}\left( 2cx\right) }\ .
\end{equation*}%
Introducing the variable $\theta =2cx,$ we come to the equation%
\begin{equation}
-\frac{d^{2}\psi }{d\theta ^{2}}+\frac{\mu \left( \mu -1\right) }{\sin
^{2}\theta }\psi =\sigma ^{2}\psi ,\ \sigma ^{2}=\frac{q^{2}}{4c^{2}},\
0\leq \theta \leq \pi ,\ \mu >1  \label{c21}
\end{equation}%
whose solutions are given by the functions%
\begin{equation}
\psi _{n}\left( \theta \right) =\sin ^{\mu }\theta \left. _{2}F_{1}\right.
\left( -n,n+2\mu ,\mu +\frac{1}{2};\frac{1}{2}\left( 1-\cos \theta \right)
\right)  \label{c22}
\end{equation}%
and the spectrum is 
\begin{equation}
\sigma _{n}^{2}=\left( n+\mu \right) ^{2},\ n=0,1,2,...  \label{c23}
\end{equation}%
We note that the hypergeometric function in Eq. (\ref{c22}) represents the
Gegenbauer polynomials $C_{n}^{\mu }\left( z\right) $ up to a numeric
factor, \cite{bev1}, Sec. 3.15, Eq. (3) (or in \cite{gr}, 8.932.1): 
\begin{equation}
C_{n}^{\mu }\left( z\right) =\frac{\Gamma \left( n+2\mu \right) }{\Gamma
\left( n+1\right) \Gamma \left( 2\mu \right) }\left. _{2}F_{1}\right. \left(
-n,n+2\mu ,\mu +\frac{1}{2},\frac{1}{2}\left( 1-z\right) \right) \ .
\label{c26}
\end{equation}

Then we can write the solutions of Eq. (\ref{c22}) as 
\begin{equation}
\psi _{n}\left( \theta \right) =\sin ^{\mu }\theta C_{n}^{\mu }\left( \cos
\theta \right) \ .  \label{c24}
\end{equation}%
For further use we recall that the Gegenbauer polynomials are solutions of
the Gegenbauer equation, \cite{bev1}, Sec. 3.15, Eq. (21) (or in \cite{gr},
8.928):%
\begin{equation}
\left( 1-z^{2}\right) y^{\prime \prime }-\left( 2\mu +1\right) zy^{\prime
}+n\left( n+2\mu \right) y=0  \label{c25}
\end{equation}%
where $n$ are non-negative integers, $y\left( z\right) =C_{n}^{\mu }\left(
z\right) $. The Gegenbauer polynomials are orthogonal on the interval $z\in %
\left[ -1,+1\right] $ with the weight $\left( 1-z^{2}\right) ^{\mu -1/2}$.
If Eq. (\ref{c26}) is used for the definition of the Gegenbauer polynomials
they are normalized by%
\begin{eqnarray}
&&\int_{-1}^{1}C_{n}^{\mu }\left( z\right) C_{m}^{\mu }\left( z\right)
\left( 1-z^{2}\right) ^{\mu -1/2}dz=  \label{c27} \\
&&\left. =\right. \int_{0}^{\pi }C_{n}^{\mu }\left( \cos \theta \right)
C_{m}^{\mu }\left( \cos \theta \right) \sin ^{2\mu }\theta d\theta =\frac{%
2^{1-2\mu }\pi \Gamma \left( n+2\mu \right) }{n!\left( n+\mu \right) \left[
\Gamma \left( \mu \right) \right] ^{2}}\delta _{nm}  \notag
\end{eqnarray}%
where $\delta _{nm}$ stands for the Kronecker delta, \cite{bev1}, Sec. 3.15,
Eqs. (16), (17).

We also will need the non-singular solutions of the equation%
\begin{equation}
\partial _{\theta }^{2}y+\left( k-1\right) \cot \theta \partial _{\theta
}y+\lambda y=0  \label{ahh.01}
\end{equation}%
which in terms of the variable $z=\cos \theta $ takes the form:%
\begin{equation*}
\left( 1-z^{2}\right) \partial _{z}^{2}y-kz\partial _{z}y+\lambda y=0\ .
\end{equation*}%
Then due to Eq. (\ref{c25}), we can wite the solutions of Eq. (\ref{ahh.01})
as%
\begin{equation}
y\left( \theta \right) =C_{n}^{\left( k-1\right) /2}\left( \cos \theta
\right) ,\ \lambda =n\left( n+k-1\right) ,\ n\geq 0\ .  \label{ahh.02}
\end{equation}

\section{Solution of the ordinary differential equations}

In this section we obtain solutions of the equations (\ref{hh.r1}), (\ref%
{hh.31}). For Eq. (\ref{hh.31}) we have 
\begin{equation}
y_{1}=e^{\pm in_{1}\theta _{1}},\ \lambda _{1}=n_{1}^{2},\ n_{1}\in \mathbb{Z%
}  \label{hh.32}
\end{equation}%
since $y_{1}$ must obey the condition of periodicity $y_{1}\left( \theta
_{1}+2\pi \right) =y_{1}\left( \theta _{1}\right) .$

For the set of the ordinary differential equations equations (\ref{hh.r1})
first we obtain solutions for $\lambda _{k-1}=0.$ In this case the equations
(\ref{hh.r1}) take the form:%
\begin{equation}
\partial _{k}^{2}y_{k}+\left( k-1\right) \cot \theta _{k}\partial
_{k}y_{k}+\lambda _{k}y_{k}=0\ .  \label{hh.41}
\end{equation}%
Due to Eqs. (\ref{ahh.01}), (\ref{ahh.02}) their solutions are given by%
\begin{equation}
y_{k}=C_{n_{k}}^{\left( k-1\right) /2}\left( \cos \theta _{k}\right) ,\
\lambda _{k}=n_{k}\left( n_{k}+k-1\right) ,\ n_{k}\geq 0\ .  \label{hh.44}
\end{equation}%
To solve the equations (\ref{hh.r1}) for $\lambda _{k-1}>0$ we present $%
y_{k} $ as follows:%
\begin{equation*}
y_{k}\left( \theta _{k}\right) =f_{k}\left( \theta _{k}\right) v_{k}\left(
\theta _{k}\right) \ .
\end{equation*}%
Then from Eq. (\ref{hh.r1}) we come to the equation for $v_{k}:$%
\begin{equation}
v_{k}^{\prime \prime }+\left( 2\frac{f_{k}^{\prime }}{f_{k}}+\left(
k-1\right) \cot \theta _{k}\right) v_{k}^{\prime }+\left( \frac{%
f_{k}^{\prime \prime }}{f_{k}}+\left( k-1\right) \cot \theta _{k}\frac{%
f_{k}^{\prime }}{f_{k}}+\lambda _{k}-\frac{\lambda _{k-1}}{\sin ^{2}\theta
_{k}}\right) v_{k}=0\ .  \label{hh.49}
\end{equation}%
To make vanish the coefficient of $v_{k}^{\prime }$ in Eq. (\ref{hh.49}) we
set%
\begin{equation*}
2\frac{f_{k}^{\prime }}{f_{k}}+\left( k-1\right) \cot \theta _{k}=0
\end{equation*}%
that gives%
\begin{equation}
f_{k}=\left( \sin \theta _{k}\right) ^{-\frac{1}{2}\left( k-1\right) }\ .
\label{hh.34}
\end{equation}%
Then for the two first terms in the coefficient of $v_{k}$ in Eq. (\ref%
{hh.49}) we get:%
\begin{equation*}
\frac{f_{k}^{\prime \prime }}{f_{k}}+\left( k-1\right) \cot \theta _{k}\frac{%
f_{k}^{\prime }}{f_{k}}=\frac{1}{2}\left( k-1\right) \left( 1-\frac{1}{2}%
\left( k-1\right) \right) \frac{1}{\sin ^{2}\theta _{k}}+\frac{1}{4}\left(
k-1\right) ^{2}\ .
\end{equation*}%
So that we come to the equation for $v_{k}:$%
\begin{eqnarray}
&&v_{k}^{\prime \prime }+\left[ \alpha _{k}^{2}-\frac{\mu _{k}\left( \mu
_{k}-1\right) }{\sin ^{2}\theta _{k}}\right] v_{k}=0,  \label{hh.50} \\
&&\mu _{k}\left( \mu _{k}-1\right) =\lambda _{k-1}-\frac{1}{2}\left(
k-1\right) \left( 1-\frac{1}{2}\left( k-1\right) \right) ,  \notag \\
&&\alpha _{k}^{2}=\frac{1}{4}\left( k-1\right) ^{2}+\lambda _{k}\ .  \notag
\end{eqnarray}%
One can see that Eq. (\ref{hh.50}) presents the Schr\"{o}dinger equation
with the symmetric P\"{o}schl-Teller potential for $\mu _{k}>1,$ Eq. (\ref%
{c21}), whose solutions are given by Eq. (\ref{c24}). We will formally use
these solutions and prove the condition $\mu _{k}>1$ during the process of
computations. For $k=2$ we have from Eq. (\ref{hh.50}):%
\begin{equation*}
\mu _{2}\left( \mu _{2}-1\right) =\left( \mu _{2}-\frac{1}{2}\right) ^{2}-%
\frac{1}{4}=\lambda _{1}-\frac{1}{4}=n_{1}^{2}-\frac{1}{4}\ .
\end{equation*}%
Then choosing a positive sign of the root we get:%
\begin{equation}
\mu _{2}=\left\vert n_{1}\right\vert +\frac{1}{2}\ .  \label{hh.33}
\end{equation}%
Here $\left\vert n_{1}\right\vert \geq 1,$ because we consider the case $%
\lambda _{1}>0,$ so that the condition $\mu _{2}>1$ is obeyed, then%
\begin{equation}
\mu _{2}=l_{1}+\frac{1}{2},\ l_{1}=\left\vert n_{1}\right\vert \ .
\label{hh.57}
\end{equation}%
Starting from $k=2$ we consequently obtain the solutions for $k\geq 3.$ For $%
\mu _{k}>1$ the solutions of Eq. (\ref{hh.50}) are given by 
\begin{equation}
v_{k}=\left( \sin \theta _{k}\right) ^{\mu _{k}}C_{n_{k}}^{\mu _{k}}\left(
\cos \theta _{k}\right) ,\ \alpha _{k}^{2}=\left( n_{k}+\mu _{k}\right)
^{2},\ n_{k}\geq 0  \label{hh.52}
\end{equation}%
with%
\begin{eqnarray}
&&\lambda _{k}=\left( n_{k}+\mu _{k}\right) ^{2}-\frac{1}{4}\left(
k-1\right) ^{2}  \label{hh.56} \\
&&\left. =\right. \left( n_{k}+\mu _{k}-\frac{1}{2}\left( k-1\right) \right)
\left( n_{k}+\mu _{k}+\frac{1}{2}\left( k-1\right) \right) \ .  \notag
\end{eqnarray}%
With the use of Eqs. (\ref{hh.50}), (\ref{hh.56}) one can write:%
\begin{equation*}
\mu _{k}\left( \mu _{k}-1\right) =\left( n_{k-1}+\mu _{k-1}\right) ^{2}-%
\frac{1}{4}
\end{equation*}%
then for $\mu _{k}$ with $k\geq 3$ one gets a recurrence relation:%
\begin{equation}
\mu _{k}=n_{k-1}+\mu _{k-1}+\frac{1}{2},\ k\geq 3  \label{hh.35}
\end{equation}%
where we choose a positive sign of the root. Repeating it $\left( k-2\right) 
$ times we get: 
\begin{equation*}
\mu _{k}=\sum_{j=2}^{k-1}n_{j}+\frac{1}{2}\left( k-2\right) +\mu
_{2}=\sum_{j=2}^{k-1}n_{j}+l_{1}+\frac{1}{2}\left( k-1\right) \ .
\end{equation*}%
Introducing the quantity%
\begin{equation}
l_{k}=l_{k-1}+n_{k}  \label{hh.36}
\end{equation}%
we can write%
\begin{equation}
\mu _{k}=l_{k-1}+\frac{1}{2}\left( k-1\right)  \label{hh.37}
\end{equation}%
then for $\lambda _{k}$ we get:%
\begin{equation}
\lambda _{k}=l_{k}\left( l_{k}+k-1\right) \ .  \label{hh.43}
\end{equation}%
We note that since for $k\geq 2$ all $n_{k}\geq 0$ then we have the
conditions for $l_{k}$ from Eq. (\ref{hh.36}):%
\begin{equation}
l_{k}\geq l_{k-1},\ 2\leq k\leq N\ .  \label{hh.38}
\end{equation}%
We also note that from Eq. (\ref{hh.43}) one gets $\lambda _{k}=0$ if $%
l_{k}=0$. The case $\lambda _{k}>0$ is realized for $l_{k}\geq 1$ that gives
the condition for $\mu _{k}$:%
\begin{equation}
\mu _{k}\geq \frac{1}{2}\left( k+1\right) >1\ \text{if}\ k\geq 2\ .
\label{hh.42}
\end{equation}%
Eq. (\ref{hh.42}) confirms appropriateness of the use of the solutions of
the quantum mechanical P\"{o}schl-Teller potential well problem.

With the use of Eqs. (\ref{hh.36}), (\ref{hh.37}) we can write $v_{k}$ in
the form:%
\begin{equation*}
v_{k}=\left( \sin \theta _{k}\right) ^{l_{k-1}+\frac{1}{2}\left( k-1\right)
}C_{l_{k}-l_{k-1}}^{l_{k-1}+\frac{1}{2}\left( k-1\right) }\left( \cos \theta
_{k}\right)
\end{equation*}%
therefore%
\begin{eqnarray}
y_{k} &=&f_{k}v_{k}=\left( \sin \theta _{k}\right)
^{l_{k-1}}C_{l_{k}-l_{k-1}}^{l_{k-1}+\frac{1}{2}\left( k-1\right) }\left(
\cos \theta _{k}\right) ,  \label{hh.46} \\
\lambda _{k} &=&l_{k}\left( l_{k}+k-1\right) ,\ l_{k}\neq 0,\ 2\leq k\leq N\
.  \notag
\end{eqnarray}%
We note that the solutions for $\lambda _{k-1}=0$ ($l_{k-1}=0$) given in Eq.
(\ref{hh.44}) have the same structure as the solutions of Eq. (\ref{hh.46}),
then it is possible to unite the solutions for all values of $l_{k}$ within
a unique formula:%
\begin{eqnarray}
y_{k} &=&\left( \sin \theta _{k}\right) ^{l_{k-1}}C_{l_{k}-l_{k-1}}^{l_{k-1}+%
\frac{1}{2}\left( k-1\right) }\left( \cos \theta _{k}\right) ,\   \label{c02}
\\
\lambda _{k} &=&l_{k}\left( l_{k}+k-1\right) ,\ l_{k}\geq l_{k-1}\geq 0,\
2\leq k\leq N\ .  \notag
\end{eqnarray}

\section{Construction of the spherical harmonics}

With the use of Eqs. (\ref{hh.32}), (\ref{c02}), and (\ref{c01}) we can
write down the N-dimensional spherical harmonics, which are the solutions of
Eq. (\ref{c13}), in the form:%
\begin{eqnarray}
&&Y_{l_{1},l_{2},...,l_{N}}\left( \theta _{1},\theta _{2},...,\theta
_{N}\right) =W_{N}\left( \theta _{1},\theta _{2},...,\theta _{N}\right)
=\prod_{k=1}^{N}y_{k}\left( \theta _{k}\right)  \label{c28} \\
&&\left. =\right. \prod_{k=2}^{N}\left( \sin \theta _{k}\right)
^{l_{k-1}}C_{l_{k}-l_{k-1}}^{l_{k-1}+\frac{1}{2}\left( k-1\right) }\left(
\cos \theta _{k}\right) e^{\pm il_{1}\theta _{1}},  \notag \\
&&\lambda _{N}=l_{N}\left( l_{N}+N-1\right) ,\ l_{N}\geq l_{N-1}\geq ...\geq
l_{2}\geq l_{1}=\left\vert n_{1}\right\vert \geq 0  \notag
\end{eqnarray}%
where $l_{N},\ l_{N-1},\ ...,\ l_{2},\ l_{1}=\left\vert n_{1}\right\vert $
are the characteristic numbers which obey the specified conditions. The form
of the spherical harmonics of Eq. (\ref{c28}) coincides with the
corresponding expressions of Refs. \cite{53be}, \cite{87h}, \cite{10a} up to
the used notations. For the sake of completeness we also compute a
normalization factor for the spherical harmonics of Eq. (\ref{c28}). To this
end we compute the integral over N-sphere $S^{N}$:%
\begin{equation*}
\left[ N_{l_{1},l_{2},...,l_{N}}\right] ^{-1/2}=%
\int_{S^{N}}Y_{l_{1},l_{2},...,l_{N}}^{\ast }Y_{l_{1},l_{2},...,l_{N}}d\Omega
\end{equation*}%
\begin{equation*}
=\int_{\theta _{N}=0}^{\pi }...\int_{\theta _{2}=0}^{\pi }\int_{\theta
_{1}=0}^{2\pi }Y_{l_{1},l_{2},...,l_{N}}^{\ast
}Y_{l_{1},l_{2},...,l_{N}}\prod_{k=1}^{N}\sin ^{k-1}\theta _{k}d\theta _{k}
\end{equation*}%
\begin{equation*}
=\prod_{k=2}^{N}\int_{\theta _{k}=0}^{\pi }\left( \sin \theta _{k}\right)
^{2l_{k-1}}\left[ C_{l_{k}-l_{k-1}}^{l_{k-1}+\frac{1}{2}\left( k-1\right)
}\left( \cos \theta _{k}\right) \right] ^{2}\sin ^{k-1}\theta _{k}d\theta
_{k}\int_{\theta _{1}=0}^{2\pi }d\theta _{1}
\end{equation*}%
\begin{equation}
=2\pi \prod_{k=2}^{N}\int_{\theta _{k}=0}^{\pi }\left[ C_{n_{k}}^{\mu
_{k}}\left( \cos \theta _{k}\right) \right] ^{2}\left( \sin \theta
_{k}\right) ^{2\mu _{k}}d\theta _{k}=2\pi \prod_{k=2}^{N}\frac{2^{1-2\mu
_{k}}\pi \Gamma \left( n_{k}+2\mu _{k}\right) }{n_{k}!\left( n_{k}+\mu
_{k}\right) \left[ \Gamma \left( \mu _{k}\right) \right] ^{2}}  \label{c29}
\end{equation}%
where Eq. (\ref{c27}) was used. We also write it in terms of the
characteristic numbers $l_{1},l_{2},...,l_{N}:$%
\begin{equation}
\left[ N_{l_{1},l_{2},...,l_{N}}\right] ^{-1/2}=\frac{1}{2}\left( 4\pi
\right) ^{N}\prod_{k=2}^{N}\frac{2^{-2l_{k-1}-k}\Gamma \left(
l_{k}+l_{k-1}+k-1\right) }{\left( l_{k}-l_{k-1}\right) !\left( l_{k}+\left(
k-1\right) /2\right) \left( \Gamma \left( l_{k-1}+\left( k-1\right)
/2\right) \right) ^{2}}\ .  \label{c30}
\end{equation}%
Finally, we remark that orthogonality of the system of the spherical
harmonics of Eq. (\ref{c28}) can be deduced from the orthogonal property of
the Gegenbauer polynomials, Eq. (\ref{c27}).

\section{Conclusions}

In this paper we present another view on the problem of construction of the
N-dimensional spherical harmonics. In our approach we effectively use the
solutions of the quantum mechanical P\"{o}schl-Teller potential well
problem. We use the eigenfunctions of the symmetric P\"{o}schl-Teller
potential well problem to construct the functional expression and the
eigenvalues for obtaining of the characteristic numbers of the spherical
harmonics. The conditions for the characteristic numbers arise naturally in
the process of solution. We expect that our approach to the problem may be
usefully applicable in consideration of problems which involve spherical
geometry of higher dimensions. For example, it can be incorporated in
schemes of construction of vector, tensor or spin spherical harmonics.
Besides that, we believe that the approach is comprehensible for students of
final years of undergraduate courses of universities. It could be utilized
in teaching of the theme of the usual two-dimensional spherical harmonics,
leastwise as complementary to the usual methods, and it allows in a simple
way to pass to higher dimensional generalizations.

\end{document}